\documentstyle{article}  

\newtheorem{D}{Definition}
\newtheorem{T}{Proposition}
\newenvironment{Th}{\paragraph{Theorem}\em }{\/\rm   
\par\vskip8pt}   
  
\def\c{{\bot}} 
  
\def\PN{\cP(\cN)} \def\cN{{\cal N}} \def\cP{{\cal P}}   
 \def\cH{{\cal H}} \def\cS{{\cal S}}   
\def\cB{{\cal B}}  
\def\cN{{\cal M}} \def\cL{{\cal L}} \def\cS{{\cal S}}   
\def\cM{{\cal M}}   
\def\cHH{${\cal H}\oplus{\cal H}$}  
\def\chh{{\cal H}\oplus{\cal H}}

\def\vagy{\vee}  
\def\es{\wedge}  

\begin{document}  
  
\title{  
{\bf On Reichenbach's common cause principle and \\  
Reichenbach's notion of common cause}}  
\author{{\it G\'abor Hofer-Szab\'o}\\  
Department of Philosophy\\  
Technical University of Budapest\\  
e-mail: gszabo@hps.elte.hu \\  
$\ $\\  
{\it Mikl\'os R\'edei}\\  
Department of History and Philosophy of Science  \\  
E\"otv\"os University, Budapest\\  
email: redei@ludens.elte.hu\\
$\ $\\ 
{\it L\'aszl\'o E. Szab\'o\/}\\  
Department of Theoretical Physics\\  
Department of History and Philosophy of Science\\  
E\"otv\"os University, Budapest\\  
e-mail: szabol@caesar.elte.hu}  
\maketitle

\abstract{It is shown that, given any finite set of pairs of random events in
a Boolean algebra which are correlated with respect to a fixed
probability measure on the algebra, the algebra can be extended in
such a way that the extension contains events that can be regarded as
common causes of the correlations in the sense of Reichenbach's
definition of common cause. It is shown, further, that, given any
quantum probability space and any set of commuting events in it which
are correlated with respect to a fixed quantum state, the quantum
probability space can be extended in such a way that the extension
contains common causes of all the selected correlations, where common
cause is again taken in the sense of Reichenbach's definition. It is argued
that these results very strongly restrict the possible ways of
disproving Reichenbach's Common Cause Principle. 
}

\section{Informal statement of the problem}  
  
Reichenbach's Common Cause Principle and the mathematically explicit
notion of common cause formulated in terms of random events and their
probabilities goes back to Reichenbach's 1956 book
(\cite{Reichenbach1956}, Section 19). Both the Common Cause Principle
and the related concept of common cause have been subjects of
investigations in a number of works, especially in the papers by
Salmon (\cite{Salmon1978}, \cite{Salmon1980}, \cite{Salmon1984}), by
Fraassen (\cite{Fraassen1977}, \cite{Fraassen1982},
\cite{Fraassen1989}), by Suppes and Zanotti (\cite{Suppes1970},
\cite{Suppes-Zanotti1981}), by Cartwright \cite{Cartwright1987} and by
Spohn \cite{Spohn1991}.  It seems that there is no general consensus
as regards the status of the Common Cause Principle and its relation
to the notion of common cause. We do not wish to evaluate here the
relative merits of the different analyses of the Principle and its
relation to the common cause notion; nor do we aim at a critical
comparison of the different suggestions as to how best to specify a
technical notion of common cause. Rather, we take the notion of common
cause as it was formulated by Reichenbach himself, and we investigate
the following problem, raised first in \cite{Redei1998}. Let
$(\cS,\mu)$ be a classical probability space, i.e. $\cS$ be a Boolean
algebra of random events and $\mu$ be a probability measure on
$\cS$. Asssume that $\{(A_i,B_i)\ \vert \ i\in I\}$ is a set of pairs
of events in $\cS$ that are statistically correlated with respect to
$\mu$: $\mu(A_iB_i)>\mu(A_i)\mu(B_i)$ $(i\in I)$, and assume, further,
that $\cS$ does not contain events $C_i$ that can be considered the
common causes of the correlation between $A_i$ and $B_i$, where common
cause is taken in the sense of Reichenbach's definition
\cite{Reichenbach1956}, Section 19 (recalled in Section \ref{cc}
below). The problem is whether the probability space $(\cS,\mu)$ can
in principle be enlarged in such a way that for each correlated pair
$(A_i,B_i)$ there exists a common cause $C_i$ of the correlation in
the larger probability space $(\cS',\mu')$. If the probability space
$(\cS,\mu)$ is such that for every pair $(A_i,B_i)$ in the set of
correlated pairs $\{(A_i,B_i)\ \vert \ i\in I\}$ the space $(\cS,\mu)$
can be enlarged in the said manner, then we say that $(\cS,\mu)$ is
{\em common cause completeable with respect to the given set of
correlations} (see the Definition \ref{defcccc} in Section \ref{scccc}
for a precise formulation). We shall prove that {\em every} classical
probability space $(\cS,\mu)$ is common cause completeable with
respect to {\em any finite} set of correlations. (Proposition
\ref{cccc} in Section \ref{scccc}). That is to say, we show that given
{\em any finite} set of correlations in a classical event structure,
one can {\em always} say that the correlations are due to some common
causes, possibly ``hidden'' ones, i.e. ones that are not part of the
inital set $\cS$ of events.
  
{\em Reichenbach's Common Cause Principle} is the claim   
that if there is a correlation between two events $A$ and   
$B$ and a direct causal connection between the correlated   
events is excluded then there exists a common cause of   
the correlation in Reichenbach's sense. We interpret   
Proposition 1 as saying that Reichenbach's Common Cause   
Principle cannot be disproved by displaying classical   
probability spaces that contain a fnite number of   
correlated events without containing a Reichenbachian   
common cause of the correlations -- the only justifiable   
conclusion one can draw is that the event structures in   
question are common cause incomplete.  
  
Statistical correlations also make sense in non-classical probability
spaces $(\cL,\phi)$, where $\cL$ is a non-distributive, orthomodular
lattice in the place of the Boolean algebra $\cS$, and where $\phi$ is
a generalized probability measure (``state'') defined on $\cL$, taking
the place of $\mu$. Such non-classical probability spaces emerge in
non-relativistic quantum mechanics and in relativistic quantum field
theory. In these theories $\cL$ is the non-Boolean, orthomodular von
Neumann lattice $\cP(\cN)$ of projections of a non-commutative von
Neumann algebra $\cN$ determined by the observables of the quantum
system. One of the difficulties in connection with interpreting
quantum theory is the alleged impossibility of existence of common
causes of the correlations in those quantum event structures. Note
that it is not quite obvious what one means by a common cause in a
quantum event structure because Reichenbach's original definition of
common cause was formulated in terms of events in a classical,
Kolmogorovian probability space, and the definition also makes
essential use of classical conditional probabilities, which do not
make sense in general in non-commutative (quantum) probability
spaces. In fact, the definition of common cause in the literature
analyzing the problem of common cause of quantum correlations is quite
different from the Reichenbachian one: typically the definition is
formulated in terms of hidden variables rather than in terms of
events, and it also makes (more or less tacitly) the extra assumption
that the hidden variables are {\em common} common causes (see
below). In this paper we wish to retain all features of Reichenbach's
original definition while applying it to quantum correlations, and we
do this by requiring explicitly the common cause to commute with the
events in the correlation. Having thus obtained a definition of common
cause in quantum event structures (see Definition \ref{defcccq1} in
Section \ref{ncc}) we define $(\cP(\cN),\phi)$ to be common cause
completeable with respect to a given set $\{(A_i,B_i)\ \vert \ i\in
I\}$ of pairs of (commuting) correlated events in $\PN$ if the
probability space $(\cP(\cN),\phi)$ can be enlarged in such a way that
each of the correlations has a common cause in the enlarged
non-classical probability space $(\cP(\cN'),\phi')$ (see the
Definition \ref{defccq} in Section \ref{ncc} for a precise
formulation). We shall prove that {\em every} $(\cP(\cN),\phi)$ is
also common cause completeable with respect to the {\em complete} set
of elements that are correlated in a given quantum state $\phi$
(Proposition \ref{cccq} in Section \ref{ncc}). Proposition \ref{cccq}
allows us to conclude that Reichenbach's Common Cause Principle cannot
be disproved even by finding quantum probability spaces that contain
correlated events without containing a Reichenbachian common cause of
the correlations; the only justifiable conclusion one can draw is that
the quantum event structures are common cause incomplete -- just like
in the classical case. This conclusion seems to be in contradiction
with the standard interpretation, according to which quantum
correlations cannot have a hidden common cause. In Section \ref{Con}
we localize the reason of this ``contradiction".  The essential point
we make is that the standard arguments showing the impossibility of
existence of a common cause of quantum correlations assume that a
common cause is a {\em common} common cause, an assumption, we claim,
is not part of Reichenbach's notion of common cause. In Section
\ref{open} we formulate a couple of open questions concerning the
notion of common cause completeablity.
  
\section{Reichenbach's notion of common cause\label{cc}}   
  
Let $({\cal S},\mu)$ be a classical probability space,   
where ${\cal S}$ is the Boolean algebra of events and   
$\mu$ is the probability measure. If the joint   
probability $\mu(A\es B)$   
of $A$ and $B$ is greater than the product of the single   
probabilities, i.e. if  
\begin{equation}\label{corr} \mu(A\es B)>\mu(A)\mu(B)   
\end{equation} then the events $A$ and $B$ are said to be   
(positively) {\em correlated}.   
  
According to Reichenbach (\cite{Reichenbach1956}, Section   
19), a probabilistic common cause type explanation of a   
correlation like (\ref{corr}) means finding an event $C$   
(common cause) that satisfies the conditions specified in   
the next definition.  
\begin{D}\label{RD} $C$ is a {\em common cause} of the   
correlation (\ref{corr}) if the following (independent)   
conditions hold:  
\begin{eqnarray} \mu(A\es B|C)&=&\mu(A|C)\mu(B|C)   
\label{off1}\\  
\mu(A\es B|C^{\bot})&=&\mu(A|C^{\bot})\mu(B|C^{\bot})   
\label{off2}\\  
\mu(A|C)&>&\mu(A|C^{\bot}) \label{nagy1}\\  
\mu(B|C)&>&\mu(B|C^{\bot})\label{nagy2}  
\end{eqnarray} where $\mu(X|Y)=\mu(X\es Y)/\mu(Y)$   
denotes the   
conditional probability of $X$ on condition $Y$, $C^\c$   
denotes the complement of $C$ and it is assumed that none   
of the probabilities $\mu(X)$, $(X=A,B,C,C^{\bot})$ is   
equal to zero.   
\end{D} We shall occasionally refer to conditions   
(\ref{off1})-(\ref{nagy2}) as ``Reichenbach(ian)   
conditions''.  
  
Reichenbach proves the following  
  
\begin{Th} Conditions (\ref{off1})-(\ref{nagy2}) imply   
(\ref{corr}); that is to say, if $A,B$ and $C$ are such   
that they satisfy conditions (\ref{off1})-(\ref{nagy2}),   
then there is a a positive correlation between $A$ and   
$B$ in the sense of (\ref{corr}).  
\end{Th}  
  
Some remarks and terminology:  
\begin{itemize}   
\item[(i)] We emphasize that, from the point of view of   
the explanatory role of $C$ as the common cause of the   
correlation, each of the independent conditions   
(\ref{off1})-(\ref{nagy2}) is equally important. For   
instance, the mere fact that an event $C$ satisfies   
(\ref{nagy1}) and (\ref{nagy2}) only, i.e., the fact that   
$C$ is statistically relevant for both $A$ and $B$, is   
not sufficient for $C$ to be accepted as an explanation   
of the correlation, since statistical relevance in and by   
itself is not sufficient to derive the correlation   
(\ref{corr}). This remains true even if, in addition to   
statistical relevance, we assume either $\mu(A\mid   
C)=\mu(B\mid C)=1$, or $C\subseteq A\es B$, since   
(\ref{off2}) can still fail, and again (\ref{corr})   
cannot be derived.

\item[(ii)] Taking either $A$ or $B$ as $C$, the four conditions
(\ref{off1})-(\ref{nagy2}) are satisfied. This means that
Reichenbach's definition accomodates the case when there is a direct
causal link between the correlated events. To put this negatively: as
it stands, Reichenbach's definition does not distinguish between
direct causal influence between the correlated events and the
correlation caused by a common cause.  Reichenbach's definition does
not exclude the following sort of common cause either:
$C\neq   
A$ and/or $C\neq B$ but  
\begin{eqnarray}  
\label{inprop1}  
(C<A \mbox{ or } C> A) \mbox{ and }\mu(C)=\mu(A)  
\end{eqnarray}  
and/or   
\begin{eqnarray}  
\label{inprop2}  
(C<B \mbox{ or } C>B) \mbox{ and }\mu(C)=\mu(B)  
\end{eqnarray}  
Such a $C$ is a common cause in the sense of the Definition 
\ref{RD} but such a $C$ should not be regarded 
as a meaningful common cause   
because $C$ is identical with  $A$ and/or   
$B$ up to a probability-0 event. If $C$ is a   
common cause such that none of (\ref{inprop1}) and   
(\ref{inprop2}) holds then we shall say   
that $C$ is a {\em proper} common cause, and in what   
follows, common cause will always mean a proper common   
cause unless stated otherwise explicitly.   
  
\item[(iii)] It can happen that, in addition to being a   
probabilistic common cause, the event $C$ logically   
implies both $A$ and $B$, i.e. $C\subseteq A\es B$. If   
this is the case then we call $C$ a {\em strong} common   
cause. If $C$ is a common cause such that $C\not\subseteq   
A$ and $C\not\subseteq B$ then $C$ is called a {\em   
genuinely probabilistic} common cause.  
  
\item[(iv)] A common cause $C$ will be called {\em   
deterministic} if   
\begin{eqnarray*}  
\mu(A|C)=&1&=\mu(B|C)\\  
\mu(A|C^{\bot})=&0&=\mu(B|C^{\bot})  
\end{eqnarray*}  
\end{itemize} Note that the notions of deterministic and   
genuinely probabilistic common cause are not negations of   
each other. There does not seem to exist any   
straightforward relation between the notions of   
deterministic, genuinely probabilistic and proper common   
causes so defined.  
  
Next we wish to determine the restrictions imposed on the   
values of the probabilities $\mu(C)$, $\mu(A\vert   
C)$, $\mu(A\vert C^\c)$, $\mu(B\vert C)$ and $\mu(B\vert C^\c)$ by the   
assumption that the correlation between $A$ and $B$ has   
$C$ as a common cause. If we assume that there exists a   
common cause $C$ in $(\cS,\mu)$ of the given correlation   
$\mu(A\es B)>\mu(A)\mu(B)$ then, using the theorem of   
total probability  
\begin{eqnarray*}  
\mu(X)=\mu(X\vert Y)\mu(Y)+\mu(X\vert Y^\c)(1-  
\mu(Y))\qquad X,Y\in\cS  
\end{eqnarray*} we can write   
\begin{eqnarray}  
\mu(A)&=&\mu(A\vert C)\mu(C)+\mu(A\vert C^\c)(1-\mu(C))  
\label{egy}\\  
\mu(B)&=&\mu(B\vert C)\mu(C)+\mu(B\vert C^\c)(1-\mu(C))  
\label{ketto}\\  
\mu(A\es B)&=&\mu(A\es B\vert C)\mu(C)+\mu(A\es B\vert   
C^\c)(1-\mu(C))  
\label{harom}\\  
&=&\mu(A\vert C)\mu(B\vert C)\mu(C) +\mu(A\vert   
C^\c)\mu(B\vert C^\c)(1-\mu(C))\label{negy}  
\end{eqnarray}   
(\ref{negy}) follows from (\ref{harom}) because of the   
screening off equations (\ref{off1})-(\ref{off2})). So   
the assumption of a common cause of the correlation   
between $A$ and $B$ implies that there exist real numbers   
$$ r_C,r_{A\vert C},r_{B\vert C},r_{A\vert   
C^\c},r_{B\vert C^\c} $$ satisfying the following   
relations  
\begin{eqnarray} 0&\leq &r_{A\vert C},r_{B\vert   
C},r_{A\vert C^\c},r_{B\vert C^\c}\leq 1\label{font4}\\  
\mu(A)&=&r_{A\vert C}r_C+r_{A\vert C^\c}(1-  
r_{C})\label{font1}\\  
\mu(B)&=&r_{B\vert C}r_C+r_{B\vert C^\c}(1-  
r_{C})\label{font2}\\  
\mu(A\es B)&=&r_{A\vert C}r_{B\vert C}r_C+r_{A^\vert   
C^\c}r_{B\vert C^\c}(1-r_C)\label{font3}\\  
0&<&r_C<1  
\label{font7}\\  
r_{A\vert C}&>&r_{A\vert C^\c}\label{font5}\\  
r_{B\vert C}&>&r_{B\vert C^\c}\label{font6}  
\end{eqnarray} Conversely, given a correlation $\mu(A\es   
B)>\mu(A)\mu(B)$ in a probability space $(\cS,\mu)$, if   
there exists an element $C$ in $\cS$ such that  
\begin{eqnarray}  
\mu(C)&=&r_C\label{t1}\\  
\mu(A\vert C)&=&r_{A\vert C}\label{t2}\\  
\mu(A\vert C^\c)&=&r_{A\vert C^\c}\label{t3}\\  
\mu(B\vert C)&=&r_{B\vert C}\label{t4}\\  
\mu(B\vert C^\c)&=&r_{B\vert C^\c}\label{t5}  
\end{eqnarray} and the numbers $r_C, r_{A\vert C},   
r_{B\vert C}, r_{A\vert C^\c}, r_{B\vert C^\c}$ satisfy   
the relations (\ref{font4})-(\ref{font6}), then the   
element $C$ is a common cause of the correlation in   
Reichenbach's sense.  
  
\begin{T} Given any correlation $\mu(A\es   
B)>\mu(A)\mu(B)$ in $(\cS,\mu)$ there exists a non-empty two   
parameter family of numbers $$ r_C(t,s), r_{A\vert   
C}(t,s), r_{B\vert C}(t,s), r_{A\vert C^\c}(t,s),   
r_{B\vert C^\c}(t,s) $$ that satisfy the relations   
(\ref{font4})-(\ref{font6}).  
\end{T}  
  
\noindent {\bf Proof}: Consider the system of 3 equations   
(\ref{font1})-(\ref{font3}) with $t=r_{A\vert C}$ and   
$s=r_{B \vert C}$ as parameters. One can then express   
$r_C, r_{A\vert C^\c}$ and $r_{B\vert C^\c}$ from   
equations (\ref{font1})-(\ref{font3}) as follows.  
\begin{eqnarray} r_C&=&  
\frac{\mu\left(A\es B\right)-\mu\left(A\right)  
\mu\left(B\right)} {\left(\mu\left(A\right)-  
t\right)\left(\mu\left(B\right)- s\right)+  
\mu\left(A\es B\right)-  
\mu\left(A\right)\mu\left(B\right)}\label{kell1}  
\\  
r_{A\vert C^\c}&=&\frac{\mu\left(A\right)-t}{   
r_C}+t=\frac{\mu\left(A\es B\right) -\mu\left(A\right) s}   
{\mu\left(B\right) -s}\label{kell2}\\  
r_{B\vert C^\c}&=&\frac{\mu\left(B\right)-s}   
{r_C}+s=\frac{\mu\left(A\es B\right) -\mu\left(B\right)   
t} {\mu\left(A\right) -t} \label{kell3}  
\end{eqnarray} Using the equations (\ref{kell1})-  
(\ref{kell3}) it is easy to verify that choosing the two   
parameters $t,s$ within the bounds  
\begin{eqnarray} 1\geq &t=r_{A\vert C}&\geq   
\frac{\mu(A\es B)}{\mu(B)}\label{par1}\\  
1\geq &s=r_{B\vert C}&\geq \frac{\mu(A\es   
B)}{\mu(A)}\label{par2}  
\end{eqnarray} the conditions (\ref{font4})-(\ref{font6})   
are satisfied.  
  
As the above proposition shows, the Reichenbach conditions allow, in
principle, for a number of different common causes, each characterized
probabilistically by the five real numbers $r_C$, $r_{A\vert C}$,
$r_{B\vert C}$, $r_{A\vert C^\c}$, $r_{B\vert C^\c}$ satisfying the
relations (\ref{font4})-(\ref{font6}). Given a correlation $\mu(A\es
B)>\mu(A)\mu(B)$, we call a set of five real numbers $r_C$, $r_{A\vert
C}$, $r_{B\vert C}$, $r_{A\vert C^\c}$, $r_{B\vert C^\c}$ {\em admissible}
if they satisfy conditions (\ref{font4})-(\ref{font6}).
  
\begin{D}\label{type} A common cause $C$ of a correlation   
$\mu(A\es B)>\mu(A)\mu(B)$ is said to have (be of) the   
{\em type} $(r_C, r_{A\vert C}, r_{B\vert C}, r_{A\vert   
C^\c}, r_{B\vert C^\c})$ if these numbers are equal to   
the probabilities indicated by the indices, i.e. if the   
equations (\ref{t1})-(\ref{t5}) hold.  
\end{D}   
  
\section{Common cause completeability -- the classical   
case\label{scccc}}  
  
Given a statistically correlated pair of events $A,B$ in   
a probability space $(\cS,\mu)$, a proper common cause   
$C$ in the sense of Reichenbach's definition does not   
necessarily exist in $\cS$. (For instance the set of   
events might contain only $I,A,B$ and their orthogonal   
complements and hence be too small to contain a proper   
common cause.) If this is the case, then we call   
$(\cS,\mu)$ {\em common cause incomplete}. The existence   
of common cause incomplete probability spaces leads to   
the question of whether such probability spaces can be   
enlarged so that the larger probability space contains a   
proper common cause of the given correlation. What is   
meant by ``enlargement'' here is contained in the   
Definition \ref{enlarge} below. Before giving this   
definition recall that the map $h\colon\cS_1\to\cS_2$   
between two Boolean algebras $\cS_1$ and $\cS_2$ is a   
{\em Boolean algebra homomorphism} if it preserves all   
lattice operations (including orthocomplementation). A   
Boolean algebra homomorphism $h$ is an {\em embedding} if   
$X\not=Y$ implies $h(X)\not=h(Y)$.  
  
\begin{D}\label{enlarge} The probability space   
$(\cS',\mu')$ is called an {\em extension} of $(\cS,\mu)$   
if there exists a Boolean algebra embedding $h$ of $\cS$   
into $\cS'$ such that  
\begin{equation}\label{mu}  
\mu(X)=\mu'(h(X)) \qquad\mbox{for all\ } X\in\cS  
\end{equation}   
\end{D}   
  
This definition, and in particular the condition   
(\ref{mu}), implies that if $(\cS'\mu')$ is an extension   
of $(\cS,\mu)$ (with respect to the embedding $h$), then   
every single correlation $\mu(A\es B)>\mu(A)\mu(B)$ in   
$(\cS,\mu)$ is carried over intact by $h$ into the   
correlation   
\begin{eqnarray*}  
\mu'(h(A)\es h(B))&=&\mu(h(A\es B))=  
\mu(A\es B)\ >\ \mu(A)\mu(B)=  
\mu'(h(A))\mu'(h(B))  
\end{eqnarray*} Hence, it makes sense to ask whether a   
correlation in $(\cS,\mu)$ has a Reichenbachian common   
cause in the extension $(\cS',\mu')$. So we stipulate  
  
\begin{D}\label{defcccctype} We say that $(\cS',\mu')$ is   
a type $(r_C, r_{A\vert C}, r_{B\vert C}, r_{A\vert   
C^\c}, r_{B\vert C^\c})$ {\em common cause completion of}   
$(\cS,\mu)$ {\em with respect to the correlated events}   
$A,B$ if $(\cS'\mu')$ is an extension of $(\cS,\mu)$, and   
there exists a Reichenbachian common cause $C\in\cS'$ of   
type $(r_C, r_{A\vert C}, r_{B\vert C}$, $r_{A\vert   
C^\c}, r_{B\vert C^\c})$ of the correlation   
$\mu'(h(A)\es h(B))>\mu'(h(A))\mu'(h(B))$.  
\end{D}  
  
\begin{D}\label{defcccc} Let $(\cS,\mu)$ be a probability   
space and $\{(A_i,B_i)\   
\vert \ i\in I\}$ be a set of pairs of correlated events   
in $\cS$. We say that $(\cS,\mu)$ is {\em common cause   
completeable with respect to the set $\{(A_i,B_i)\ \vert   
\ i\in I\}$ of correlated events} if, given any set of   
admissible numbers $(r^i_C, r^i_{A\vert C}, r^i_{B\vert   
C}, r^i_{A\vert C^\c}, r^i_{B\vert C^\c})$ for every   
$i\in I$, there exists a probability space $(\cS',\mu')$   
such that for every $i\in I$ the space $(\cS',\mu')$ is a   
type $(r^i_C, r_{A\vert C}, r^i_{B\vert C}, r^i_{A\vert   
C^\c}, r^i_{B\vert C^\c})$ common cause extension of   
$(\cS,\mu)$ with respect to the correlated events   
$A_i,B_i$.  
\end{D}  
  
\begin{T}\label{cccc} Every classical probability space   
$(\cS,\mu)$ is common cause completeable with respect to   
any {\em finite} set of correlated events.   
\end{T}  
  
\bigskip   
\noindent {\bf Proof}: Let $\{(A_i,B_i)\ \vert \   
i=1,2\ldots N\}$ be a finite number of correlated pairs.   
We prove the statement by induction on the index $i$. We   
prove first that given the single pair $(A_1,B_1)=(A,B)$   
of correlated events in $(\cS,\mu)$ and any admissible   
numbers $(r_C, r_{A\vert C}, r_{B\vert C}, r_{A\vert   
C^\c}, r_{B\vert C^\c})$ there exists a type $(r_C,   
r_{A\vert C}, r_{B\vert C}, r_{A\vert C^\c}, r_{B\vert   
C^\c})$ common cause completion of $(\cS,\mu)$ with   
respect to $(A,B)$. We prove this statement in two steps.   
In Step 1 we construct an extension $(\cS',\mu')$ of   
$(\cS,\mu)$. In Step 2 we show that given any admissible   
numbers, the probability measure $\mu'$ can be chosen in   
such a way that there exists a proper common cause in   
$\cS'$ that has the type specified by the admissible   
numbers. Finally, we shall argue that if $(\cS^{n-  
1},\mu^{n-1})$ is a common cause completion of   
$(\cS,\mu)$ with respect to the set of $n-1$ correlations   
between $A_i$ and $B_i$ $(i=1,\ldots n-1)$, then there   
exists a common cause completion of $(\cS,\mu)$ with   
respect to the $n$ correlations.   
  
\paragraph*{Step~1}
By Stone's theorem we may assume   
without loss of generality that $\cS$ is a field of   
subsets of a set $\Omega$. Let $\Omega_1$ and $\Omega_2$   
be two identical copies of $\Omega$, distinguishable by   
the indices 1 and 2, and let $\cS_1$ and $\cS_2$ be the   
corresponding two coopies of $\cS$:  
\begin{eqnarray*}  
\Omega_i&\equiv&\{(x,i)\quad\vert\quad   
x\in\Omega\}\qquad\qquad\qquad\qquad   
\qquad i=1,2\\  
\cS_i&\equiv&\{\{(x,i)\quad\vert\quad x\in   
X\}\quad\vert\quad X\in\cS\}  
\qquad\qquad i=1,2  
\end{eqnarray*} Let $h_i$ $(i=1,2)$ denote the Boolean   
algebra isomorphisms between $\cS$ and $\cS_i$ $(i=1,2)$:  
\begin{eqnarray*}  
\cS\ni X\mapsto h_i(X)=\{(x,i)\quad\vert\quad x\in X\}   
\qquad\qquad i=1,2  
\end{eqnarray*}   
Furthermore, let $\cS'$ be the set of subsets of   
$\Omega_1\cup\Omega_2$ having the form $h_1(X)\cup   
h_2(Y)$, i.e.  
\begin{eqnarray*}  
\cS'\equiv\{h_1(X)\cup h_2(Y)\quad\vert\quad X,Y\in\cS\}  
\end{eqnarray*}   
We claim that $\cS'$ is a Boolean algebra of subsets of   
$\Omega_1\cup\Omega_2$ with respect to the usual set   
theoretical operations $\cup,\cap,\bot$ and that the map   
$h$ defined by  
\begin{equation} h(X)\equiv h_1(X)\cup h_2(X) \hskip .5cm   
X\in\cS  
\end{equation} is an embedding of $\cS$ into $\cS'$. To   
see that $\cS'$ is a Boolean algebra one only has to show   
that $\cS'$ is closed with respect to the set theoretical   
operations of join, meet and complement, and this is a   
straightforward consequence of the fact that $\cS$,   
itself being a Boolean algebra with respect to the set   
theoretical operations, is closed with respect to these   
operations. Checking the homomorphism properties of $h$   
is a routine task.   
  
We now define a measure $\mu'$ on $\cS'$ that has the   
property (\ref{mu}). Let $r_i$ $(i=1,2,3,4)$ be arbitrary   
four real numbers in the interval $\lbrack 0,1\rbrack$.   
One can define a $\mu'$ measure on $\cS'$ by  
\begin{eqnarray*} \label{muc} \mu'(h_1(X)\cup   
h_2(Y))&\equiv &r_1\mu(X\cap(A\cap B)) +   
r_2\mu(X\cap(A\cap B^\c))\\  
& &+r_3\mu(X\cap(A^\c\cap B)) + r_4\mu(X\cap(A^\c\cap   
B^\c))\\  
& &+(1- r_1)\mu(Y\cap(A\cap B)) + (1-r_2)\mu(Y\cap(A\cap   
B^\c))\\  
& &+(1- r_3)\mu(Y\cap(A^\c\cap B)) + (1-  
r_4)\mu(Y\cap(A^\c\cap B^\c))   
\end{eqnarray*}   
Since $A\cap B,A\cap B^\c,A^\c\cap B$ and $A^\c\cap B^\c$   
are disjoint and their union is $\Omega$ it follows that   
\begin{eqnarray*}   
\mu'(h_1(X)\cup h_2(X))=\mu'(h(X))=\mu(X) \qquad\qquad   
X\in\cS   
\end{eqnarray*}   
Hence $(\cS'\mu')$ is indeed an extension of the original   
probability space $(\cS,\mu)$.   
  
\paragraph*{Step~2}
Choose any value of the parameters $t,s$ within the bounds specified
by (\ref{par1})- (\ref{par2}), and consider the corresponding numbers
$r_{A\vert C}=t$, $r_{B \vert C}=s$ and $r_C,r_{A\vert C^\c},r_{B\vert
C^\c}$, the latter ones defined by (\ref{kell1})-(\ref{kell3}). We
claim that the probability space $(\cS',\mu')$ constructed in Step 1
is a common cause extension of $(\cS,\mu)$ of type $(r_C, r_{A\vert
C}, r_{B\vert C}, r_{A\vert C^\c}, r_{B\vert C^\c})$ with respect to
the correlation between $A$ and $B$, if the numbers $r_i$
$(i=1,2,3,4)$ defining $\mu'$ by the formula (\ref{muc}) are given by
\begin{eqnarray*}   
r_1&=&\frac{r_Cr_{A\vert C}(1-r_{B\vert C})}{\mu(A)-  
\mu(A\es B)}\\  
r_2&=&\frac{r_Cr_{A\vert C}r_{B\vert C}}{\mu(A\es B)}\\  
r_3&=&\frac{r_Cr_{B\vert C}(1-r_{A\vert C})}{\mu(B)-  
\mu(A\es B)}\\  
r_4&=&\frac{r_C(1-r_{A\vert C}-r_{B\vert C}+r_{A\vert   
C}r_{B\vert C})} {\mu(A^\c\es B^\c)}  
\end{eqnarray*}   
To show that $(\cS',\mu')$ is a common cause completion   
of $(\cS,\mu)$ one only has to display a proper common   
cause $C$ in $\cS'$ of the correlation. We claim that   
$C=h_1(\Omega)\cup h_2(\emptyset)$ is a proper common   
cause. Clearly, $C$ is a {\em proper} common cause if it   
is a common cause. To see that $C$ is a common cause   
indeed one can check by explicit calculation that the   
following hold   
\begin{eqnarray}  
\mu'(h_1(\Omega)\cup h_2(\emptyset))&=&r_C\\  
\mu'((h_1(A)\cup h_2(A))\vert (h_1(\Omega)\cup   
h_2(\emptyset))&=&r_{A\vert C}\\  
\mu'((h_1(B)\cup h_2(B))\vert (h_1(\Omega)\cup   
h_2(\emptyset))&=&r_{B\vert C}\\  
\mu'((h_1(A)\cup h_2(A))\vert \lbrack h_1(\Omega)\cup   
h_2(\emptyset)  
\rbrack^\c)&=&r_{A\vert C^\c}\\  
\mu'((h_1(B)\cup h_2(B))\vert \lbrack h_1(\Omega)\cup   
h_2(\emptyset)  
\rbrack^\c)&=&r_{B\vert C^\c}  
\end{eqnarray} Since the numbers $r_{A\vert C}, r_{B   
\vert C}, r_C, r_{A\vert C^\c}, r_{B\vert C^\c}$ were   
chosen so that they satisfy the conditions (\ref{font4})-  
(\ref{font6}), $C$ is indeed a common cause.  
  
Assume now that there exists a common cause completion   
$(\cS^{n-1},\mu^{n-1})$ of $(\cS,\mu)$ that contains a   
common cause $C_i$ of each correlation $\mu(A_i\es   
B_i)>\mu(A_i)\mu(B_i)$ $(i=1,\ldots n-1)$. Consider the   
correlation between $A_n$ and $B_n$. By repeating the two   
steps (Step 1-Step 2) one can construct a common cause   
completion $(\cS^n,\mu^n)$ of $(\cS^{n-1},\mu^{n-1})$   
that contains a common cause $C_n$ of the correlation   
between $A_n$ and $B_n$. To complete the induction one   
only has to see that $(\cS^n,\mu^n)$ also contains common   
causes of each of the correlations between   
$h_n(A_i),h_n(B_i)$ $(i=1,\ldots n-1)$, where $h_n$ is   
the Boolean algebra embedding of $\cS^{n-1}$ into   
$\cS^n$. But $h_n(C_i)$ $(i=1,2\ldots )$ are clearly   
common causes of the correlations between   
$h_n(A_i),h_n(B_i)$ $(i=1,\ldots n-1)$ because $h_n$ is a   
homomorphism preserving $\mu_{n-1}$.   
  
\section{Common cause completeability -- the quantum   
case\label{ncc}}   
  
Statistical correlations also make sense in non-classical   
probability structures $(\cL,\phi)$, where $\cL$ is a   
non-distributive lattice of events and $\phi$ is a   
generalized probability measure (``state'') on $\cL$.   
Such non-classical probability spaces arise in quantum   
theory, where $\cL$ is the non-distributive, orthomodular   
lattice of projections $\cP(\cN)$ of a non-commutative   
von Neumann algebra $\cN$ determined by the set of   
observables of a quantum system (for the operator   
algebraic notions used here without definition see eg.   
\cite{Kadison-Ringrose1983},   
\cite{Kadison-Ringrose1986}.) A map   
$\phi\colon\cP(\cN)\to\lbrack 0,1\rbrack$ on such an   
event structure is called a {\em state} if it is additive   
on orthogonal projections in the following sense:  
\begin{eqnarray*}  
\phi(\vee_i P_i)   
=\sum_i\phi(P_i)\qquad\qquad P_i\bot P_j   
\quad i\not=j   
\end{eqnarray*} A positive, linear functional on a von   
Neumann algebra $\cN$ is called {\em normal} state if its   
restriction to the lattice of projections is a state   
(i.e. it is additive) in the above sense. The restriction   
of a normal state to a Boolean sublattice of $\PN$ is a   
classical probability measure, so normal states are the   
analogues of classical probability measures. If $\cM$   
acts on the Hilbert space $\cH$, then a normal state is   
always of the form $\phi(X)=Tr(W X)$ with some density   
matrix $W$. We call a pair $(\PN,\phi)$ with a normal   
state $\phi$ a {\em quantum probability space}. Two {\em   
commuting} events $A,B$ in a quantum probability space   
$(\PN,\phi)$ are called (positively) correlated if  
\begin{eqnarray*}\label{qcorr}  
\phi(A\es B)>\phi(A)\phi(B)   
\end{eqnarray*} Given a correlation in a quantum   
probability space, we may want to ask if there is a   
Reichenbachian common cause in $\PN$ of the correlation.   
By a Reichenbachian common cause we mean a $C\in\cP(\cM)$   
which commutes with both $A$ and $B$ and satisfies the   
Reichenbachian conditions (\ref{off1})-(\ref{nagy2}). To   
be explicit we stipulate the following  
  
\begin{D} \label{defcccq1} The event $C\in\PN$ is a   
common cause of the correlation (\ref{qcorr}) between two   
commuting events $A,B$ in a quantum probability space   
$(\cP(\cM),\phi)$ if   
\begin{enumerate}  
\item $C$ commutes with both $A$ and $B$;  
\item the following four conditions (analogous to   
(\ref{off1})-(\ref{nagy2})) are satisfied  
\end{enumerate}  
\begin{eqnarray*} {\phi(A\es B\es C)\over\phi(C)}&=&   
{\phi(A\es C)\over\phi(C)}{\phi(B\es C)\over\phi(C)}  
\label{qoff1}\\  
{\phi(A\es B\es C^\c)\over\phi(C^\c)}&=& {\phi(A\es   
C^\c)\over\phi(C^\c)}{\phi(B\es C^\c)  
\over\phi(C^\c)}\label{qoff2}\\  
{\phi(A\es C)\over\phi(C)}&>&{\phi(A\es   
C^\c)\over\phi(C^\c)}  
\label{qnagyobb1}\\  
{\phi(B\es C)\over\phi(C)}&>&{\phi(B\es   
C^\c)\over\phi(C^\c)}\label{qnagyobb2}  
\end{eqnarray*}   
Similarly to the classical case, a common cause $C$ is called {\em
proper} if it differs from both $A$ and $B$ by more than a
$\phi$-measure zero event.
\end{D}  
  
Having this definition, we can define the type of the   
common cause in a quantum probability space exactly the   
same way as in the classical case, and we can also speak   
of admissible numbers etc. Just like a classical   
probability space, a quantum probability space   
$(\PN,\phi)$ may contain a correlation without containing   
a proper common cause of the correlation in the sense of   
Definition \ref{defcccq1}. If this is the case, then we   
call the quantum probability space {\em common cause   
incomplete}, and we may ask if the quantum probability   
space can be enlarged so that the enlarged space contains   
a proper common cause. What is meant by ``enlargement''   
is specified in the next definition, which is completely   
analogous to the Definition \ref{enlarge}.   
  
\begin{D}\label{qenlarge} The quantum probability space   
$(\cP(\cN'),\phi')$ is an {\em extension} of the quantum   
probability space $(\PN,\phi)$ if there exists an   
embedding $h$ of $\PN$ into $\cP(\cN')$ such that  
\begin{eqnarray*}  
\phi'(h(X))=\phi(X) \qquad\mbox{for all\ } X\in\PN  
\end{eqnarray*}   
\end{D}   
By an embedding is meant here a lattice homomorphism that   
preserves all lattice operations (including the   
orthocomplementaion) and such that $X\not=Y$ implies   
$h(X)\not=h(Y)$.  
\begin{D}\label{defccqtype} We say that   
$(\cP(\cN'),\phi')$ is a type $(r_C, r_{A\vert C},   
r_{B\vert C}, r_{A\vert C^\c}, r_{B\vert C^\c})$ {\em   
common cause completion of} $(\PN,\phi)$ {\em with   
respect to the correlated events} $A,B$ if   
$(\cP(\cN'),\phi')$ is an extension of $(\PN,\phi)$, and   
there exists a Reichenbachian common cause   
$C\in\cP(\cN')$ of type $(r_C$, $r_{A\vert C}$, $r_{B\vert   
C}$, $r_{A\vert C^\c}$, $r_{B\vert C^\c})$ of the correlation   
$\phi'(h(A)\es h(B))>\phi'(h(A))\phi'(h(B))$.  
\end{D} We can now give the definition of common cause   
competeability in the quantum case:  
  
\begin{D}\label{defccq} Let $(\PN,\mu)$ be a quantum   
probability space and $\{(A_i,B_i)\ \vert \ i\in I\}$ be   
a set of pairs of correlated events in $\PN$. We say that   
$(\PN,\phi)$ is {\em common cause completeable with   
respect to the set $\{(A_i,B_i)\ \vert   
\ i\in I\}$ of correlated events} if, given any set of   
admissible numbers $(r^i_C, r^i_{A\vert C}, r^i_{B\vert   
C}, r^i_{A\vert C^\c}, r^i_{B\vert C^\c})$ for every   
$i\in I$, there exists a quantum probability space   
$(\cP(\cN'),\phi')$ such that for every $i\in I$ the   
space $(\cP(\cN'),\phi')$ is a type $(r^i_C, r_{A\vert   
C}, r^i_{B\vert C}, r^i_{A\vert C^\c}, r^i_{B\vert   
C^\c})$ common cause extension of $(\PN,\phi)$ with   
respect to the correlated pair $(A_i,B_i)$.  
\end{D}  
  
\begin{T}\label{cccq} Every quantum probability space   
$(\PN,\phi)$ is common cause completeable with respect to   
the set of pairs of events that are correlated in the   
state $\phi$.   
\end{T}   
\bigskip \noindent {\bf Proof}: The proof is divided into   
two parts. In the first part we construct an extension of   
the quantum probability space $(\PN,\phi)$, this is done   
in two steps. In Step 1 the quantum probability space   
$(\PN,\phi)$ is embedded into the quantum probability   
space $(\cP(\cH\oplus\cH),\phi_2)$ with a suitable state   
$\phi_2$ extending $\phi$, where $\cH$ is the Hilbert   
space the von Neumann algebra $\cN$ is acting on. In Step   
2 this latter quantum probability space is embedded into   
$(\cP(\cH'),\phi')$, where $\cH'$ is a Hilbert space   
constructed explicitly. We show in the second part of the   
proof that for {\em any} correlated pair $(A,B)$ in   
$(\PN,\phi)$ and for {\em any} admissible set of numbers   
there exists in $(\cP(\cH'),\phi')$ a Reichenbachian   
common cause of type defined by the admissible numbers.  
  
\noindent {\bf Step 1} We may assume without loss of   
generality that $\cN$ is acting on a Hilbert space $\cH$.   
Let $\cH\oplus\cH$ be the direct sum of $\cH$ with itself   
and consider the map $h_2$ defined by   
\begin{eqnarray*}  
\PN\ni X\mapsto h_2(X)\equiv X\oplus   
X\in\cP(\cH\oplus\cH)  
\end{eqnarray*}   
Then $h_2$ is an embedding of $\PN$ as an orthomodular   
lattice into the orthomodular lattice   
$\cP(\cH\oplus\cH)$. Let $\phi_2$ be a state  defined on   
$\cP(\cH\oplus\cH)$ by the density matrix   
$\frac{1}{2}W\oplus\frac{1}{2}W$, where $W$ is the   
density matrix belonging to $\phi$. Clearly, $\phi_2$ has   
the property  
\begin{eqnarray*}  
\phi_2(h_2(X))=\phi(X)\qquad\qquad X\in\PN  
\end{eqnarray*}   
So the probability space $(\cP(\cH\oplus\cH),\phi_2)$ is   
an extension of $(\PN,\phi)$. Since every density matrix   
is a convex combination of (possibly countably infinite   
number of) one dimensional projections, there exist   
vectors $\psi_k\in(\cH\oplus\cH)$ and non-negative   
numbers $\lambda_k$ ($k=1\ldots$) such that   
$\sum_k^{\infty}\lambda_k=1$ and   
\begin{eqnarray*}\label{lambda}  
\frac{1}{2}W\oplus\frac{1}{2}W=\sum_k^{\infty}\lambda_k   
P_{\psi_k}  
\end{eqnarray*} (Here, and in what follows, $P_{\xi}$   
denotes the projection to the one dimensional subspace   
spanned by the Hilbert space vector $\xi$.)  
  
\noindent {\bf Step 2} Let $\cH'$ be the set of functions $g\colon
N\mapsto\cH\oplus\cH$ such that $sup_n \| g(n)\|_2<\infty$, where
$\|\xi\|_2$ is the norm of $\xi\in\cal H\oplus\cH$. Then the set
$\cH'$ is a complex linear space with the pointwise operations
($(\kappa_1 g_1 + \kappa_2 g_2)(n)
=\kappa_1g_1(n)+\kappa_2g_2(n)$). It is elementary to check that
$\cH'$ becomes a Hilbert space with the scalar product
$\langle,\rangle'$ defined by
\begin{eqnarray*}  
\langle   
g_1,g_2\rangle'\equiv\sum_{k=1}^{\infty}\lambda_k\langle   
g_1(k),g_2(k)\rangle_2  
\end{eqnarray*} where $\langle,\rangle_2$ is the scalar   
product in $\cH\oplus\cH$ and the numbers $\lambda_k$ are   
taken from (\ref{lambda}). The map   
$h'\colon\cB(\cH\oplus\cH)\to\cB(\cH')$ defined by  
\begin{eqnarray*} (h'(Q))g(n)=Q(g(n))\qquad\quad n\in N,   
\ g\in\cH'  
\end{eqnarray*} is an algebra homomorphism from the   
algebra $\cB(\cH\oplus\cH)$ of all bounded operators on   
$\cH\oplus\cH$ into the algebra $\cB(\cH')$ of all   
bounded operators on $\cH'$; furthermore,   
$h(Q_1)\not=h_2(Q_2)$ if $Q_1\not=Q_2$, and a routine   
reasoning shows that if $Q_n\in\cB(\cH\oplus\cH)$ is a   
sequence of operators such that for some   
$Q\in\cB(\cH\oplus\cH)$ we have $Q_n\xi\to Q\xi$ for all   
$\xi\in\cH\oplus\cH$, then $(h'(Q_n))g\to h'(Q)g$ for all   
$g\in \cH'$. This means that $h'$ is continuous in the   
respective strong operator topologies in   
$\cB(\cH\oplus\cH)$ and $\cB(\cH')$. It follows that if   
$A,B$ are two projections in $\cP(\cH\oplus\cH)$, then  
\begin{eqnarray*} h'(A\es B)&=&h'(s-\lim (AB)^n)  
=s-\lim (h'(AB)^n)=s-\lim(h'(A)h'(B))^n\\  
&=&h'(A)\es h'(B)  
\end{eqnarray*} It follows then that the restriction of   
$h'$ to $\cP(\cH\oplus\cH)$ is an embedding of   
$\cP(\cH\oplus\cH)$ as an orthomodular lattice into the   
orthomodular lattice $\cP(\cH')$. Let $\xi_l$   
($l=1,\ldots dim(\cH\oplus\cH)$) be an orthonormal basis   
in \cHH. Then the elements $g_{kl}\in\cH'$ defined by  
\begin{eqnarray*} g_{kl}(n)&=&\left\{\begin{array}{ll}  
\delta_{nk}\frac{1}{\sqrt{\lambda_k}}\xi_l&\mbox{ if   
$\lambda_k\not=0$}\\  
0 &\mbox{ if $\lambda_k=0$}  
\end{array}  
\right.  
\end{eqnarray*} form an orthonormal basis in $\cH'$.   
($\delta_{nk}$ denoting the Kronecker symbol.) The linear   
operator $W'$ on $\cH'$ defined by  
\begin{eqnarray*} (W'g)(n)\equiv\lambda_n P_{\psi_n}g(n)   
\qquad n\in N,\  g\in\cH'  
\end{eqnarray*} is easily seen to be a density matrix,   
hence it defines a state $\phi'$ on $\cP(\cH')$. For   
$A\in\cP(\chh)$ we have  
\begin{eqnarray*} \phi'(h'(A))&=&Tr(W'h(A))  
=\sum_{k=1}^{\infty}\sum_{l=1}^{dim(\chh)} \langle   
g_{kl},W'h(A)g_{kl}\rangle'\\  
&=&\sum_{k=1}^{\infty}\sum_{l=1}^{dim(\chh)}\sum_{n=1}^{  
\infty}  
\lambda_n\langle g_{kl}(n),\big\lbrack   
W'h(A)g_{kl}\big\rbrack(n)\rangle_2\\  
&=&  
\sum_{k=1}^{\infty}\sum_{l=1}^{dim(\chh)}\sum_{n=1}^{  
\infty}  
\lambda_n\langle\delta_{nk}\frac{1}{\sqrt{\lambda_k}}   
\xi_l \lambda_n   
P_{\psi_n}A\frac{1}{\sqrt{\lambda_k}}\xi_l\rangle_2\\  
&=&  
\sum_{k=1}^{\infty}\sum_{l=1}^{dim(\chh)}  
\lambda_k\langle\xi_l,P_{\psi_k}A\xi_l\rangle_2\\  
&=&  
\sum_{l=1}^{dim(\chh)}\langle\xi_l,\big\lbrack   
\sum_{k=1}^{\infty}  
\lambda_kP_{\psi_k}\big\rbrack\xi_l\rangle_2  
= Tr(WA)=\phi(A)  
\end{eqnarray*} So $(\cP(\cH'),\phi')$ is an extension of   
the quantum probability space $(\cP(\chh),\phi_2)$. It   
follows that $(\cP(\cH'),\phi')$ is an extension of the   
probability space $(\cP(\cN),\phi)$, where the embedding   
$h\colon\PN\to\cP(\cH')$ is given by $h'\circ h_2$.  
  
We now claim that for any given pair of events $(A,B)$ in   
$\PN$ that are correlated with respect to $\phi$ and for   
any given set of admissible numbers $(r_C, r_{A\vert C},   
r_{B\vert C}, r_{A\vert C^\c}, r_{B\vert C^\c})$ the   
probability space $(\cP(\cH'),\phi')$ constructed above   
contains a proper Reichenbachian common cause $C$ of type   
$(r_C, r_{A\vert C}, r_{B\vert C}, r_{A\vert C^\c},   
r_{B\vert C^\c})$ of the correlation.   
  
Indeed, given the numbers $(r_C, r_{A\vert C}, r_{B\vert   
C}, r_{A\vert C^\c}, r_{B\vert C^\c})$, the event $C$   
defined below by (\ref{c}) is a common cause of type   
$(r_C, r_{A\vert C}, r_{B\vert C}, r_{A\vert C^\c},   
r_{B\vert C^\c})$.   
  
\begin{equation} \label{c} C=P_{\alpha}\vagy   
P_{\beta}\vagy P_{\gamma}\vagy P_{\delta}  
\end{equation} where $P_{\alpha}$, $P_{\beta}$,   
$P_{\gamma}$ and $P_{\delta}$ are projections in   
$\cP(\cH')$ defined by  
\begin{eqnarray*}   
\begin{array}{rcl}  
(P_{\alpha}g)(k)&=&P_{\alpha_k}g(k)\\  
(P_{\beta}g)(k)&=&P_{\beta_k}g(k)\\  
(P_{\gamma}g)(k)&=&P_{\gamma_k}g(k)\\  
(P_{\delta}g)(k)&=&P_{\delta_k}g(k)  
\end{array}  
\qquad\qquad k\in N, g\in\cH'  
\end{eqnarray*} where $P_{\alpha_k}$, $P_{\beta_k}$,   
$P_{\gamma_k}$ and $P_{\delta_k}$ are projections in \cHH   
defined by  
\begin{eqnarray*}   
\alpha_k&=& \left\{ \begin{array}{ll} 0& \mbox{if   
$\langle\psi_k,(A\wedge B^{\bot}\oplus A\wedge   
B^{\bot})\psi_k\rangle_2=0 $}\\  
\cos \omega^{\alpha}_{k}\alpha^1_k + \sin  
\omega^{\alpha}_{k}\alpha^2_k & \mbox{if   
$\langle\psi_k,(A\wedge B^{\bot}\oplus A\wedge   
B^{\bot})\psi_k\rangle_2\neq 0 $}   
\end{array}   
\right.\label{zdef1}\\  
\end{eqnarray*} where  
\begin{eqnarray*}   
\cos^2 \omega^{\alpha}_{k} &=& r_{A|Z}(1 -   
r_{B|Z})r_{Z}\frac{\langle\psi_k,(A\wedge B^{\bot}\oplus   
A\wedge B^{\bot})\psi_k\rangle_2}{\phi(A^{\bot}\wedge   
B)}\label{zdef2}\\  
\alpha^1_k&=& \frac{1}{\langle\psi_k,(A\wedge   
B^{\bot}\oplus A\wedge B^{\bot})\psi_k\rangle_2}(A\wedge   
B^{\bot}\oplus A\wedge B^{\bot})\psi_k \label{zdef3}\\  
\alpha^1_k &\bot& \alpha^2_k \in A\wedge B^{\bot}\oplus   
A\wedge B^{\bot}\\  
\left\langle \alpha^2_k,   
\alpha^2_k\right\rangle_2&=&1   
\end{eqnarray*}   
\begin{eqnarray*}   
\beta_k&=&\left\{ \begin{array}{ll} 0 & \mbox{if   
$\langle\psi_k,(A\wedge B \oplus A\wedge   
B)\psi_k\rangle_2=0 $}\\  
\cos \omega^{\beta}_{k} \beta^1_k +   
\sin \omega^{\beta}_{k}\beta^2_k & \mbox{if   
$\langle\psi_k,(A\wedge B \oplus A\wedge   
B)\psi_k\rangle_2\neq 0 $}   
\end{array}   
\right.\label{zdef4}   
\end{eqnarray*} where   
\begin{eqnarray*}   
\cos^2 \omega^{\beta}_{k} &=&   
r_{A|Z}r_{B|Z}r_C\frac{\langle\psi_k,(A\wedge B \oplus   
A\wedge B)\psi_k\rangle_2}{\phi(A\wedge   
B)}\label{zdef5}\\  
\beta^1_k&=&\frac{1}{\langle\psi_k,(A\wedge B \oplus   
A\wedge B)\psi_k\rangle_2}(A\wedge B \oplus A\wedge   
B)\psi_k   
\label{zdef6}\\  
\beta^1_k &\bot& \beta^2_k \in A\wedge B\oplus A\wedge   
B\\  
\left\langle \beta^2_k,   
\beta^2_k\right\rangle_2&=&1   
\end{eqnarray*}   
\begin{eqnarray*}   
\gamma_k&=&\left\{ \begin{array}{ll} 0 & \mbox{if   
$\langle\psi_k,(A^{\bot}\wedge B\oplus A^{\bot}\wedge   
B)\psi_k\rangle_2=0 $}\\  
\cos \omega^{\gamma}_{k}\gamma^1_k + \sin   
\omega^{\gamma}_{k}\gamma^2_k & \mbox{if   
$\langle\psi_k,(A^{\bot}\wedge B\oplus A^{\bot}\wedge   
B)\psi_k\rangle_2\neq 0 $}   
\end{array}   
\right.\label{zdef7}   
\end{eqnarray*} where   
\begin{eqnarray*}   
\cos^2 \omega^{\gamma}_{k} &=& r_{B|Z}(1 -   
r_{A|Z})r_{Z}\frac{\langle\psi_k,(A^{\bot}\wedge B\oplus   
A^{\bot}\wedge B)\psi_k\rangle_2}{\phi(A^{\bot}\wedge   
B)}\label{zdef8}\\  
\gamma^1_k&=&\frac{1}{\langle\psi_k,( A^{\bot}\wedge   
B\oplus A^{\bot}\wedge B)\psi_k\rangle_2}(A^{\bot}\wedge   
B\oplus A^{\bot}\wedge B)\psi_k   
\label{zdef9}\\  
\gamma^1_k &\bot& \gamma^2_k \in A^{\bot}\wedge B\oplus   
A^{\bot}\wedge B\\  
\left\langle \gamma^2_k,   
\gamma^2_k\right\rangle'&=&1   
\end{eqnarray*}   
\begin{eqnarray*}   
\delta_k&=&\left\{ \begin{array}{ll} 0 & \mbox{if   
$\langle\psi_k,(A^{\bot}\wedge B^{\bot}\oplus   
A^{\bot}\wedge B^{\bot}\psi_k\rangle_2) = 0 $}\\  
\cos \omega^{\delta}_{k} \delta^1_k + \sin   
\omega^{\delta}_{k}\delta^2_k & \mbox{if   
$\langle\psi_k,(A^{\bot}\wedge B^{\bot}\oplus   
A^{\bot}\wedge B^{\bot})\psi_k\rangle_2 \neq 0 $}   
\end{array}   
\right.\label{zdef10}   
\end{eqnarray*} where   
\begin{eqnarray*}   
\cos^2 \omega^{\delta}_{k} &=& r_Z(1-r_{A|Z} - r_{B|Z} +   
r_{A|Z}r_{B|Z})\frac{\langle\psi_k,(A^{\bot}\wedge   
B^{\bot}\oplus A^{\bot}\wedge   
B^{\bot})\psi_k\rangle_2}{\phi(A^{\bot}\wedge   
B^{\bot})}\label{zdef11}\\  
\delta^1_k&=&\frac{1}{\langle\psi_k,(A^{\bot}   
\wedge B^{\bot}\oplus A^{\bot}\wedge   
B^{\bot})\psi_k\rangle_2}(A^{\bot}\wedge B^{\bot}\oplus   
A^{\bot}\wedge B^{\bot})\psi_k   
\label{zdef12}\\  
\delta^1_k &\bot& \delta^2_k \in A^{\bot}\wedge   
B^{\bot}\oplus A^{\bot}\wedge B^{\bot}\\  
\left\langle \delta^2_k,   
\delta^2_k\right\rangle_2&=&1   
\end{eqnarray*}   
  
Since $\alpha^1_k, \beta^1_k, \gamma^1_k$ and   
$\delta^1_k$ are unit vectors, $\alpha_k$, $\beta_k$,   
$\gamma_k$, $\delta_k$ also are unit vectors in \cHH. The   
element $C$ commutes with $h(A)$ and $h(B)$ because   
\begin{eqnarray*}   
P_{\alpha_k} &<& A\wedge B^{\bot}   
\oplus A\wedge B^{\bot}\\  
P_{\beta_k} &<& A\wedge B \oplus A\wedge B\\  
P_{\gamma_k} &<& A^{\bot}\wedge B \oplus A^{\bot}\wedge   
B\\  
P_{\delta_k} &<& A^{\bot}\wedge B^{\bot} \oplus   
A^{\bot}\wedge B^{\bot}   
\end{eqnarray*}   
  
So to show that $C$ is indeed a common cause of the said   
type we just have to show that the following hold  
\begin{eqnarray}  
\phi'(C)&=&r_C\label{ezt0}\\  
\frac{\phi'(h'(A)\es C)}{\phi'(C)}&=&r_{A\vert   
C}\label{ezt}\\  
\frac{\phi'(h'(B)\es C)}{\phi'(C)}&=&r_{B\vert   
C}\label{ezt2} \\  
\frac{\phi'(h'(A)\es C^\c)}{\phi'(C^\c)}&=&r_{A\vert   
C^\c} \label{ezt3}\\  
\frac{\phi'(h'(B)\es C^\c)}{\phi'(C^\c)}&=&r_{B\vert   
C^\c}\label{ezt4}  
\end{eqnarray} We show (\ref{ezt})-(\ref{ezt4}) by   
showing first that the following hold  
\begin{eqnarray}  
\phi'(h'(A)\es C)&=&r_{A\vert C}r_C\label{elobb1}\\  
\phi'(h'(B)\es C)&=&r_{B\vert C}r_C\label{elobb2} \\  
\phi'(h'(A)\es C^\c)&=&r_{A\vert   
C^\c}r_{C^\c}\label{elobb3}\\  
\phi'(h'(B)\es C^\c)&=&r_{B\vert   
C^\c}r_{C^\c}\label{elobb4}  
\end{eqnarray} which imply (\ref{ezt})-(\ref{ezt4}) if   
$\phi'(C)=r_C$.   
  
We compute $\phi'(h'(A)\es C)$ first.   
\begin{eqnarray*}  
\phi'(h(A)\es C)&=&\phi'((h'(A)\es (P_{\alpha}\vagy   
P_{\beta})=\phi'(P_{\alpha}\vagy P_{\beta})=  
\phi'(P_{\alpha})+\phi'(P_{\beta})  
\end{eqnarray*} where we have used the fact that   
$P_{\alpha}$ $P_{\beta}$ $P_{\gamma}$ and $P_{\delta}$   
are pairwise orthogonal projections. We can compute   
$\phi'(P_{\alpha})$ as follows  
\begin{eqnarray}  
&&\phi'(P_{\alpha})=Tr(W'P_{\alpha})=  
\sum_{k=1}^{\infty}  
\sum_{l=1}^{dim(\chh)} \langle   
g_{kl},W'P_{\alpha}g_{kl}\rangle'  \nonumber \\  
& &=\sum_{k=1}^{\infty}\sum_{l=1}^{dim(\chh)}\sum_{n=1}^{   
\infty}\lambda_n  
\langle\delta_{nk}\frac{1}{\sqrt{\lambda_k}}\xi_l,  
\lambda_nP_{\psi_n}P_{\alpha_n}\delta_{nk}  
\frac{1}{\sqrt{\lambda_k}}\xi_l\rangle_2  \nonumber \\  
& &=\sum_{k=1}^{\infty}\sum_{l=1}^{dim(\chh)}\lambda_k   
\langle\xi_l, P_{\psi_k}P_{\alpha_k}\xi_l\rangle_2=Tr_2   
(P_{\psi_k}P_{ \alpha_k})  \nonumber \\  
& &=\sum_{k=1}^{\infty}\lambda_k\vert\langle   
\psi_k,\alpha_k \rangle\vert^2=  
\sum_{k=1}^{\infty}\lambda_k\vert\langle\psi_k,\cos   
\omega^{\alpha}_k\alpha^1_k+  
\sin\omega^{\alpha}_k\alpha^2_k\rangle\vert^2\nonumber\\  
& &=\sum_{k=1}^{\infty}\lambda_k\big\lbrack\cos   
\omega^{\alpha }_k\langle\psi_k,  
\alpha^1_k\rangle+\sin\omega^{\alpha}_k\langle\psi_k,   
\alpha^2_k\rangle \big\rbrack^2\label{csillag}  
\end{eqnarray} The second term in (\ref{csillag}) is   
equal to zero because $\alpha^1_k\bot\alpha^2_k$ by   
definition, and in view of the definition of $\alpha^1_k$   
and $\alpha^2_k$ we can also write  
\begin{eqnarray*}   
0=\langle\alpha^1_k,\alpha^2_k\rangle_2&=&\langle\big   
\lbrack A\es B^\c\oplus A\es   
B^\c\big\rbrack\psi_k,\alpha^2_k\rangle_2\\  
&=&\langle\psi_k, \big\lbrack A\es B^\c\oplus A\es   
B^\c\big\rbrack\alpha^2_k\rangle_2=  
\langle\psi_k,\alpha^2_k\rangle_2  
\end{eqnarray*} Since we have   
\begin{eqnarray*}  
\langle\psi_k,\alpha^1_k\rangle_2=1  
\end{eqnarray*} it follows that  
\begin{eqnarray*}  
&&\phi'(P_{\alpha}))=\sum_{k=1}^{\infty}\lambda_k  
\cos^2\omega^{\alpha}_k\\  
& &=\sum^{\infty}_{k=1}r_{A\vert C}(1-r_{B\vert C})r_C   
\frac{\langle\psi_k,(A\es B^\c\oplus A\es   
B^\c)\psi_k\rangle_2} {\phi(A\es B^\c)}\\  
& &=r_{A\vert C}(1-r_{B\vert C})r_C\frac{1}{ \phi(A\es   
B^\c)}\sum^{\infty}_{k=1}\lambda_k\langle\psi_k,(A\es   
B^\c\oplus A\es B^\c)\psi_k\rangle_2\\  
& &=r_{A\vert C}(1-r_{B\vert C})r_C\frac{1}{\phi(A\es   
B^\c)}\phi(A\es B^\c)=  
r_{A\vert C}(1-r_{B\vert C})r_C  
\end{eqnarray*} In a completely analogous way one obtains   
\begin{eqnarray*}  
\phi'(P_{\beta}))=r_{A\vert C}r_{B\vert C}r_C  
\end{eqnarray*} And so   
\begin{eqnarray*}  
\phi'(P_{\alpha})+\phi'(P_{\beta})=r_{A\vert C}r_C  
\end{eqnarray*} which is (\ref{elobb1}). A similar   
derivation shows that  
\begin{eqnarray*}  
\phi'(h(B)\es C)=r_{B\vert C}r_C  
\end{eqnarray*} which is (\ref{elobb2}). Since $C$   
commutes with both $h(A)$ and $h(B)$ we can write  
\begin{eqnarray*}  
\phi'(h(A)\es C^\c)&=&\phi'(h(A))-\phi'(h(A)\es   
C)=\phi(A)-r_{A\vert C}r_C=r_{A\vert C^\c}r_{C^\c}\\  
\phi'(h(B)\es C^\c)&=&\phi'(h(B))-\phi'(h(B)\es   
C)=\phi(B)-r_{B\vert C}r_C=r_{B\vert C^\c}r_{C^\c}  
\end{eqnarray*} which establishes (\ref{elobb3}) and   
(\ref{elobb4}). One can compute $\phi'(P_{\gamma})$ and   
$\phi'(P_{\delta})$ exactly the same way as   
$\phi'(P_{\alpha})$ and $\phi'(P_{\beta})$, and one   
obtains  
\begin{eqnarray*}  
\phi'(C)&=&\phi'(P_{\alpha}\vagy P_{\beta}\vagy   
P_{\gamma}\vagy P_{\delta})=  
\phi'(P_{\alpha})+\phi'(P_{\beta})+\phi'(P_{\gamma})+\phi   
'(P_{\delta})\\  
&=&r_{A\vert C}(1-r_{B\vert C})r_{C}+r_{A\vert   
C}r_{B\vert C}r_C+  
r_{B\vert C}(1-r_{A\vert C})r_C \\&+&r_C(1-r_{A\vert C}-  
r_{B\vert C}+r_{A\vert C}r_{B\vert C})=r_C  
\end{eqnarray*} which shows (\ref{ezt0}), which, together   
with (\ref{elobb1})-(\ref{elobb4}) proves 
(\ref{ezt0})-(\ref{ezt4}).   
  
\section{Reichenbach's Common Cause Principle and common   
cause completeability\label{Con}}  
  
{\em Reichenbach's Common Cause Principle} is a non-trivial
metaphysical claim about the causal structure of the physical world:
if a direct causal influence between the probabilisticaly correlated
events $A$ and $B$ is not possible, then there exists a (proper)
common cause of the correlation (in Reichenbach's sense). There exist
both classical and quantum probability spaces that contain
correlations without containing proper common causes of the
correlations; hence, if one wants to maintain Reichenbach's Common
Cause Principle, one must be able to claim that there might exist
``hidden'' events (``hidden'' in the sense of not being accounted for
in the common cause incomplete event structure) that can be
interpreted as the common causes of the correlations. If such
``hidden'' common cause events exist, then there must exist an
extension of the original probability space, an extension that
accommodates the common causes. Propositions \ref{cccc} and \ref{cccq}
tell us that such extensions are always possible. In other words,
Propositions \ref{cccc} and \ref{cccq} show that a necessary condition
for a common cause explanation of correlations in both classical and
quantum event structures can always be satisfied. To put this
negatively: one cannot disprove Reichenbach's Common Cause Principle
by proving that the necessary condition (common casue completeability)
for its validity cannot be satisfied.

It is generally accepted that the Reichenbach conditions are just
necessary conditions. If an event $C$ must satisfy also some
Supplementary Conditions (in addition to the Reichenbachian
conditions) to qualify as a common cause, then a disproof of
Reichenbach's Common Cause Principle requires establishing that there
exists no event whatsoever that satisfies {\em both} the
Reichenbachian conditions {\em and} the Supplementary Conditions. It
goes without saying that such a disproof requires first the
specification of the Supplementary Conditions. Propositions \ref{cccc}
and \ref{cccq} impose strong restrictions on the possible mathematical
specifications of the Supplementary Conditions: these conditions
cannot be formulated in terms of the probabilities $p(C)$, $p(A|C)$,
$p(B|C)$, $p(A|C\-^{\bot})$ and $p(B|C\-^{\bot})$. This is because the
assumptions in Propositions \ref{cccc} and \ref{cccq} contain no
restrictions whatsoever on these probabilities -- beyond the
Reichenbach conditions.  Therefore, the hypothetical Supplementary
Conditions should be specified entirely in terms of the
algebraic/logical structure of events.
  
The conclusion that probability spaces are typically common cause
completeable seems to contradict the received view concerning
correlations predicted by quantum mechanics and quantum field
theory. According to the standard interpretation an explanation of the
quantum correlations by assuming a direct causal influence between the
correlated quantum events is excluded by the theory of relativity, and
there cannot exist Reichenbachian common causes of the correlations
either because, as the argument goes, the assumption of a common cause
of correlations in Reichenbach's sense implies Bell's inequality (this
view is present in \cite{Fraassen1989}, \cite{Skyrms1984}
\cite{Butterfield1989} and \cite{Spohn1991} and it has made its way into
textbooks already (see \cite{textbook})).

But there is no contradiction here at all because the present paper's
analysis differs from the standard interpretation. To see where the
differences are, let us recall in the present paper's terminology
and notation the standard argument in favor of the claim ``existence
of Reichenbachian common causes of correlations implies Bell's
inequality''. Consider four pairs
\begin{eqnarray*}  
(A_1,B_2);(A_1,B_2);(A_2,B_1);(A_2,B_2)  
\end{eqnarray*}  
of commuting events in $\PN$ that are   
correlated in the state $\phi$. Assume that there exists   
a {\em common} common cause $C$ of the four correlations;   
i.e. assume that there exists a single $C\in\PN$ that is   
a common cause (in the sense of the Definition  
\ref{defccq}) of {\em all} four correlated pairs.  
  
Using the notation $\phi(X\vert Y)=\frac{\phi(X\es   
Y)}{\phi(Y)}$ for commuting $X,Y\in\PN$ we can write then  
  
\begin{eqnarray}  
\label{}  
\begin{array}{rcl}  
\phi(A_i\es B_j\vert C)&=&\phi(A_i\vert C)\phi(B_j\vert   
C)\\  
\phi(A_i\es B_j\vert C^\c)&=&\phi(A_i\vert   
C^\c)\phi(B_j\vert C^\c)\\  
\phi(A_i\es B_j)&=&\phi(A_i\es B_j\vert C)\phi(C) +   
\phi(A_i\es B_j\vert C^\c)\phi(C^\c) \\  
&=&\phi(A_i\vert C)\phi(B_j\vert C)\phi(C) +  
\phi(A_i\vert C^\c)\phi(B_j\vert C^\c)\phi(C^\c)  
\end{array}\ \ i,j=1,2\label{B3}  
\end{eqnarray}   
The elementary inequality for   
numbers $a_i,b_j\in\lbrack 0,1\rbrack$ $(i,j=1,2)$  
\begin{equation}\label{real}  
\vert a_ib_i+a_ib_j+a_jb_i-a_jb_j\vert\leq 2  
\end{equation} implies   
\begin{eqnarray}\label{B}  
\vert \phi(A_1\vert C)\phi(B_1\vert C)+\phi(A_1\vert   
C)\phi(B_2\vert C)  
+\phi(A_2\vert C)\phi(B_1\vert C)-\phi(A_2\vert   
C)\phi(B_2\vert C)\vert  \leq 2  
\end{eqnarray}   
Multiplying (\ref{B}) by $\phi(C)$ and by   
$\phi(C^\c)$, adding the two resulting inequalities and   
using (\ref{B3}) we obtain  
\begin{equation}\label{Bell}  
\vert \phi(A_1\es B_1)+\phi)(A_1\es B_2)+\phi(A_2\es   
B_2)-\phi(A_2\es B_2)\vert \leq 2  
\end{equation} The inequality (\ref{Bell}) is known as   
Bell's inequality, and it is known not to hold for every   
quantum state $\phi$ that predicts correlation between   
four projections (see eg. \cite{Summers1990a} and   
\cite{Summers1990b}).  
  
The crucial assumption in the above derivation of the inequality
(\ref{Bell}) is that $C$ is a common cause {\em for all four}
correlated pairs; i.e. that $C$ is a {\em common} common cause, shared
by the different correlations. Without this assumption Bell's
inequality {\em cannot} be derived. But there does not seem to be any
obvious reason why common causes should also be common common causes,
whether of quantum or of any other sort of correlations. In our
interpretation of Reichenbach's notion of common cause there is
nothing that would justify such an assumption; hence if such an
assumption is made, it needs extra support.  It should be mentioned
that while the impossibility of (non-probabilistic) {\it common}
common causes of the (non-probabilistic) GHZ correlations has been
proved in the paper \cite{Belnap-Szabo1996}, it remains open in that
paper whether non-common common causes of the GHZ correlations exist.

\section{Open questions\label{open}}  
  
To decide whether a particular event structure is common cause
incomplete does not seem to be a trivial matter. In a previous paper
the problem was raised whether the event structure defined by
(algebraic relativistic) quantum field theory is common cause
incomplete, and this problem is still open (see \cite{Redei1997} and
\cite{Redei1998} for a precise formulation of the question). It is
even conceivable that the explicitly formulated axioms that define
algebraic quantum field theory -- and thereby the set of all events --
are not strong enough to decide the issue, which, if true, would be
especially interesting.
  
It also is an open mathematical question whether one can have at all
{\em common cause closed probability spaces}, where ``common cause
closedness'' of a probability space means that for every pair of
correlated events there exists in that probability space a proper
common cause of the correlation in Reichenbach's sense. It is
important that common cause is meant a {\em proper\/} common
cause. This qualification on non-triviality is necessary for it is not
difficult to show (using standard tensor product procedures) that
every quantum probability space can be enlarged in such a way that the
enlarged quantum probability space is common cause closed in the
improper, formal sense that for every correlated pair $(A,B)$ there
exists at least one $C\leq A, C\not=A$ such that $\phi(C)=\phi(A)$ and
such a $C$ satisfies the Reichenbach conditions.
  
Whether or not common cause closed probability spaces exist, it is not
reasonable to expect a probabilistic physical theory to be common
cause closed. This is because one does not expect to have a proper
common cause explanation of probabilistic correlations that arise as a
consequence of a direct physical influence between the correlated
events, or which are due to some logical relations between the
correlated events. One would want to have a common cause explanation
of correlations only between events that are neither directly causally
related, nor do they stand in straightforward logical consequence
relations to each other. Thus a precise notion of causal
(in)dependence, different from the notion of the standard
probabilistic independence (correlation) is needed.  Perhaps the
notion of ``logical independence" (see the refs.
\cite{95a},\cite{95b} and \cite{Redei1998}) can be useful here. Two
orthocomplemented sub-lattices $\cL_1$ and $\cL_2$ of an orthomodular
lattice $\cL$ are called {\em logically independent} if $A\es B\not=0$
for any $A\in\cL_1$ and $B\in\cL_2$. This is an independence condition
that obtains between spacelike separated local systems in the sense of
(algebraic) quantum field theory; so this logical independence
condition can be viewed as a lattice theoretic formulation of
``separatedness'' of certain events. It seems reasonable then to
expect a probabilistic physical theory $(\cL,\mu)$ to be common cause
closed with respect to the correlated elements in every two, logically
independent, commuting sublattices $\cL_1,\cL_2$. It is not known if
this is possible at all.
  
As we have argued, quantum correlations cannot have a {\em common}
common cause in general. This raises the question of whether classical
correlations exist that cannot have a {\em common} common cause. Note
that Bell's inequality is never violated in classical probability
theory, so one cannot obtain an answer to this question by referring
to the violation of Bell's inequality.  What satisfaction of the Bell
inequalities does guarante is the existence of a multi-valued
($2^{n}$-valued, in case of $n$ events) hidden parameter theory, while
the Reichenbachian {\it common} common cause is equivalent to a
two-valued hidden parameter theory. In this sense the behavior of
correlations with respect to Bell's inequality is not a good indicator
of (non)existence of {\em common} common causes. (That Bell's
inequality is not a good indicator of (non)existence of common causes,
is clear from Propositions \ref{cccc} and \ref{cccq}.) There exist
{\em classical\/} probability spaces containing different pairs of
events that are correlated with respect to a fixed probability measure
and which cannot have a common common cause. (The proof of this
assertion will be published elswhere.)

\section*{Acknowledgement}

Work supported by AKP, by OTKA (contract
numbers: T 025841 and F 023447) and by the Dibner Institute MIT, where
M. R\'edei was staying during the 1997/98 academic year as a Resident
Fellow. We also wish to thank the audiences -- especially N. Belnap,
R. Clifton and W. Salmon -- of two seminars held in October 1997 at the
Center for Philosophy of Science, Pittsburgh University.

\end{document}